\documentclass[aps,prb,showpacs,reprint,superscriptaddress]{revtex4-2}

\usepackage{graphicx}
\usepackage{color}
\usepackage{amsmath}
\usepackage{bbm}
\usepackage{amssymb}
\usepackage{nameref}
\usepackage{hyperref}
\usepackage{siunitx} 

\usepackage{dsfont}

\usepackage{graphicx}
\usepackage{capt-of}

\usepackage[utf8]{inputenc}	
\usepackage[T1]{fontenc}
\usepackage{bm}

\usepackage[svgnames]{xcolor}

\input epsf

\begin{document}

\title{Topological properties and gap structure of the paired state in twisted TMD bilayers within a two-band effective model}


\author{Palash Saha}
\affiliation{Academic Centre for Materials and Nanotechnology, AGH University of Krakow, Al. Mickiewicza 30, 30-059 Krakow, Poland}
\author{Micha{\l} Zegrodnik}%
 \email{michal.zegrodnik@agh.edu.pl}
\affiliation{Academic Centre for Materials and Nanotechnology, AGH University of Krakow, Al. Mickiewicza 30, 30-059 Krakow, Poland}%

\date{\today}

\begin{abstract}
We present a theoretical study motivated by the recent experimental results which demonstrate superconductivity emerging from flat topological bands of twisted transition metal dichalcogenide (TMD) bilayers. To capture the non-trivial band topology of the system, we employ an effective two-band Kane-Mele-like model and substitute it with Coulomb repulsion and inter-site pairing interactions treated at the Hartree-Fock mean-field level. Assuming a real-space pairing scenario, we analyze the resulting superconducting gap symmetry and the stability of the paired state as a function of band filling and applied displacement field. Finally, we highlight the interplay between superconductivity and non-trivial topology, detailing how interaction-induced effects and the evolution of the density of states shape the resulting superconducting phase diagram.

\end{abstract}

\maketitle

\section{\label{sec:level1}Introduction}

The discovery of flat electronic bands in moir\'e superlattices has established twisted van der Waals materials as highly tunable platforms for exploring strongly correlated quantum matter \cite{het1,het2,het3,het4,het5,het6,het7,AndreiMacDonald2020,m_review}. While early efforts were heavily focused on magic-angle twisted bilayer graphene \cite{cao2018correlated,cao2018unconventional}, the field has rapidly expanded to transition metal dichalcogenide (TMD) moir\'e systems \cite{Xia2026, m_review,Zhao2024,Li2021,Huang2021}. Unlike graphene, twisted TMD homobilayers do not require fine-tuning to a specific "magic" angle to achieve flat bands. Furthermore, they are characterized by strong spin-orbit coupling, massive Dirac fermions, and coupled spin-valley degrees of freedom, which collectively give rise to a rich landscape of interaction-driven phenomena, including Mott insulators \cite{sc_w,Ghiotto2021}, generalized Wigner crystals \cite{mo2,Liu_2026}, and topologically nontrivial states \cite{Mohammad2026,jin2026observationmottquantumspin,fc1,fc2,fc3}.

Among the family of moir\'e materials, twisted homobilayers WSe$_2$ (tWSe$_2$) and MoTe$_2$ (tMoTe$_2$) have recently garnered intense experimental and theoretical interest due to the direct observation of unconventional superconductivity \cite{Xia2025,Xia2026,Guo2025,Guo2026,exp_sc,sun2026twistangleevolutionvalleypolarizedfractional}. A key advantage shared by both systems is that the out-of-plane displacement field serves as a powerful tuning knob that changes the interlayer potential difference. This allows to drive the phase transitions between the paired state and other correlated and/or topological states $in$ $situ$. In both systems superconductivity is observed close to half-filling in relatively narrow range of displacement fields in the proximity of the Mott insulating state \cite{Xia2026,sun2026twistangleevolutionvalleypolarizedfractional}. Additionally, in tMoTe$_2$ for different twist angles the paired state has been observed also for low values of the displacement field in close proximity of the fractional quantum anomalous Hall insulator (FQAH). In spite of some difference between the two systems, both of them are characterized by the same symmetry of the crystal structure and large spin-orbit splitting at the $K$ points of the Brillouin zone, leading to similar effective models which differ only by slight adjustment of the parameters such as hopping and interaction strength \cite{Devakul2021,wan_1,wan_2,WeiCheng2026}. In both cases, the precise nature of the observed paired state and whether the pairing is mediated primarily by purely electronic correlations, electron-phonon coupling, or a synergistic interplay of both remains a subject of ongoing debate.

Motivated by the experimental observations of the paired state in the twisted TMD homobilayers we 
analyze the principal features of the superconducting state within an effective two-band model composed of two moir\'e orbitals localized at the MX and XM stacking sites, where M corresponds to Transition Metal (like e.g. Mo or W) and X corresponds to Chalcogen (e.g., Te or Se). Additionally, we supplement the model with onsite Coulomb repulsion term and intersite pairing term which can originate from a kinetic exchange-based mechanism. To determine the principal features of the paired state, we apply a self-consistent Hartree-Fock (HF) approximation and focus on the evolution of the SC pairing symmetries arising purely due to intersite exchange interactions as a function of both band-filling and displacement field which correspond to the two experimentally controllable parameters. We analyze the singlet and triplet contributions to the paired state together with the inter- and intra- orbital gap amplitudes. Additionally, we supplement our paper with the analysis of topological features of the resultant SC phase. Within our approach we obtain a superconducting dome
residing in the low-displacement field regime. According to our analysis the obtained stability regime of the paired state is determined mostly by the evolution of the van Hove singularity across the phase diagram and the paired state has a topological character. Interestingly, by changing the displacement field one can induce a topological phase transition between two paired states characterized by a different value of the Chern number.

The paper is structured as follows:  In Sec. II, we introduce the effective two-band model and outline the implementation of the HF approach. Sec. III explores the system's pairing in the presence of intersite exchange interaction followed by the gap symmetry analysis. At the end, the analysis of the topological features of the paired phase is put forward. Finally, in Section. IV we summarize and conclude our results.

\begin{figure}[t]
\centering
\includegraphics[width=0.95\linewidth, height=3.8cm]{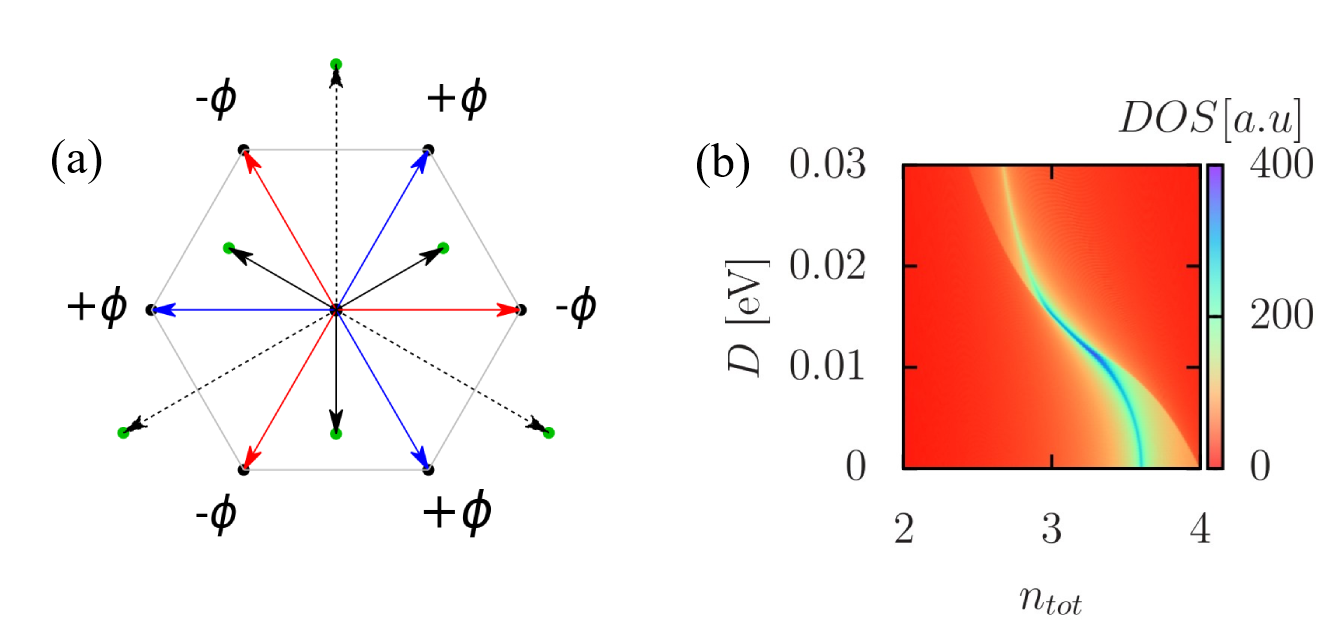}
\caption{(a) The sign of the direction dependent complex phase for the spin-up hoppings ($\sigma_z=1$) to the six NNN neighbors [cf. Eq. (\ref{ham1})]. The remaining hopping amplitudes shown in the figure are real. The green and black dots represent the lattice sites corresponding to MX and XM, stacking configurations, respectively. The position of Van Hove singularity in (b) influences the behavior of the superconducting gap away from half-filling as shown in the following.}
\label{dos}
\end{figure}

\section{Model And Method}

The electronic structure of a twisted TMD homobilayer near the valence band edge can be described in terms of localized Wannier orbitals centered at high-symmetry stacking regions of the moir\'{e} unit cell. The effective model presented in Ref. \cite{wan_1} contains three orbitals per unit cell corresponding to the MM, MX, and XM stacking configurations, where M and X correspond to metal and chalcogen atoms, respectively. The MX and XM moir\'e orbitals form a honeycomb lattice while the MM orbital corresponds to an interpenetrating triangular lattice. This configuration gives rise to three bands whose orbital character varies with the twist angle and the strength of interlayer tunneling.  It has been shown that for small to intermediate twist angles \cite{wan_1} the MM orbital degree of freedom can be integrated out in an adiabatic approximation since it is separated by an energy offset $\delta \sim 20$--$30\,\mathrm{meV}$ from the remaining two and does not directly participate in the low-energy dynamics. Instead, its effect is encoded through effective hopping processes between the MX and XM states. Additionally, the intra-sublattice MX and XM hoppings are complex with both direction and spin dependent complex phase. The resulting effective description reduces to a honeycomb lattice model analogous to the Kane-Mele model. The obtained two-band tight-binding Hamiltonian has the following form
\begin{equation}
\begin{aligned}
\mathcal{H}_t =\;&
\sum_{\langle ij \rangle \sigma} t_1 \, \hat{c}_{i1\sigma}^\dagger \hat{c}_{j2\sigma}
+ \sum_{\langle\langle ij \rangle\rangle l\sigma} t_2 \, e^{i \phi \nu_{ij} \sigma^z} \hat{c}_{il\sigma}^\dagger \hat{c}_{jl\sigma} \\
&+ \sum_{\langle\langle\langle ij \rangle\rangle\rangle \sigma} t_3 \, \hat{c}_{i1\sigma}^\dagger \hat{c}_{j2\sigma}
+ D \sum_{i \sigma} (\hat{n}_{i1\sigma}-   \hat{n}_{i2\sigma}),
\end{aligned}
\label{ham1}
\end{equation}
where $\hat{c}_{il\sigma}^\dagger$ ($\hat{c}_{il\sigma}$) creates (annihilates) a hole at the site $i$ of the sublattice $l$ (where $l=1,2$) with spin $\sigma$ (where $\sigma^z = \pm 1$ for $\sigma = \uparrow, \downarrow$) and $\hat{n}_{il\sigma}=\hat{c}_{il\sigma}^\dagger\hat{c}_{il\sigma}$ is the number operator on a lattice site. The two triangular sublattices corresponding to $l=1,2$ with MX and XM moir\'e orbitals, respectively, together form a honeycomb lattice. 
\begin{figure}[t]
\centering
\includegraphics[width=1.0\linewidth, height=8.4cm]{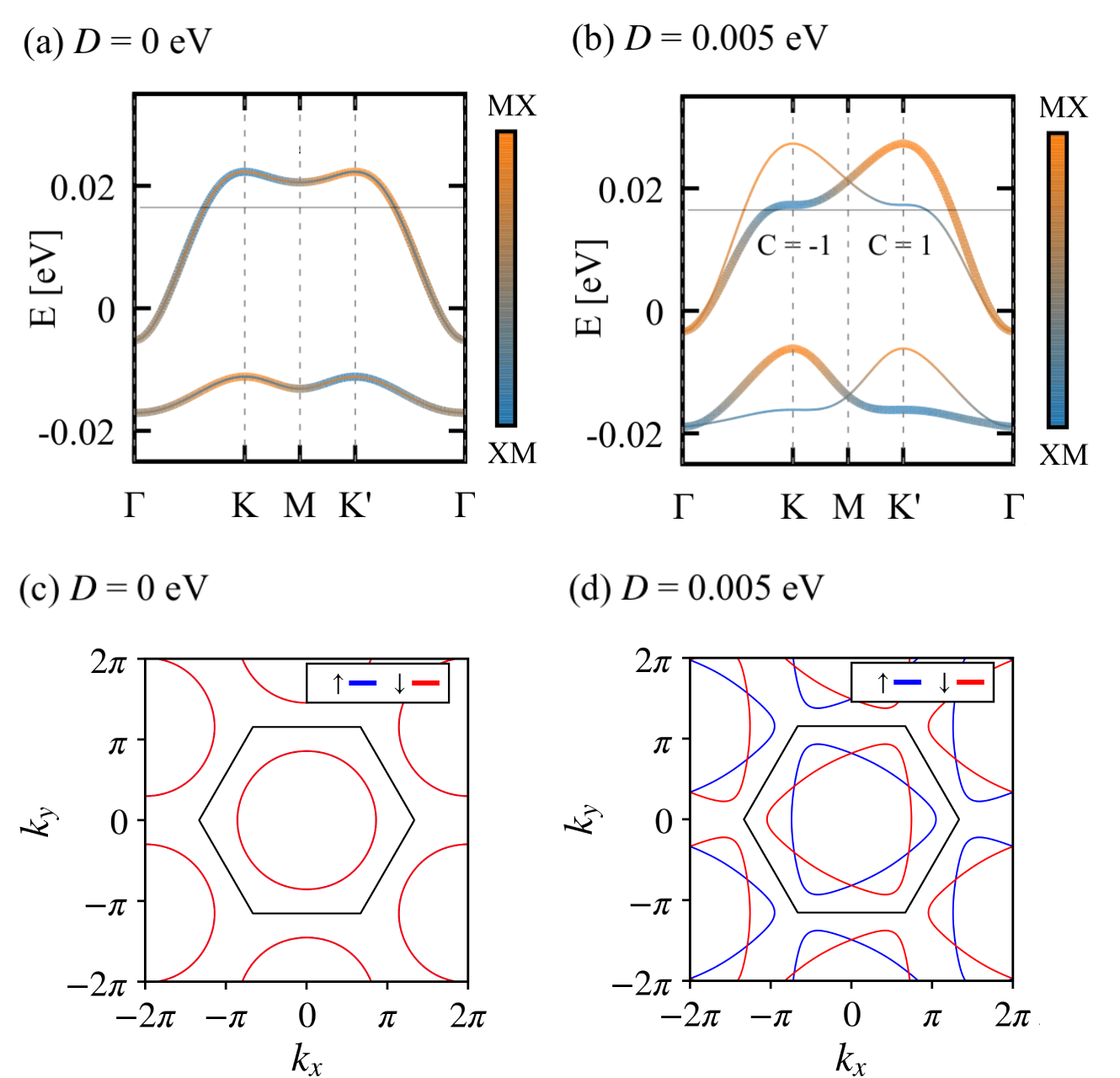}
\caption{
Dispersion relations and Fermi surfaces of the effective two-band non-interacting model.
Panels (a) and (b) show the band dispersions for two selected values of the displacement field, while panels (c) and (d) show the corresponding Fermi surfaces at half filling.
The blue and red curves denote the two spin sectors.
In the dispersion plots, the thicker (thinner) curves correspond to the spin-up (spin-down) sector. For non-zero displacement field, single particle bands are topological with the two spin sectors carrying opposite value of the Chern number ($C=\pm 1$).
}
\label{FS1}
\end{figure}
Summations over $\langle ij \rangle$, $\langle\langle ij \rangle\rangle$, and $\langle\langle\langle ij \rangle\rangle\rangle$ correspond to nearest neighbor (NN), next-nearest neighbor (NNN) and second nearest-neighbor (NNNN) hoppings on the honeycomb lattice. Therefore, the first and the third is of interorbital (intersublattice) character, while the remaining one is of the intraorbital (intrasublattice) character. The sign factor ($\nu_{ij}=\pm 1$) is determined by the direction of the intra-orbital hopping, as indicated in Fig.~\ref{dos}(a). It should be noted that from the experimental perspective so far the paired state has been identified approximately in the twist angle range $3.6^{\circ}\lesssim\theta\lesssim5.0^{\circ}$ for tWSe$_2$ \cite{Xia2026,Guo2026} and for $\theta=3.83^{\circ}$ and $\theta=5.78^{\circ}$ for the case of tMoTe$_2$ \cite{exp_sc,sun2026twistangleevolutionvalleypolarizedfractional}. In the following we select $\theta=3.8$ and set the values of hopping parameters to $t_1$ = 5.69 meV and $t_2$ = $t_3$ = 3.72 meV based on the analysis provided in Ref. \cite{wan_1}.

\begin{table}[t]
\caption{There are six discrete values assigned to the symmetry index \( M \), each associated with a specific parity value \( p \), as shown in the second column. 
The third and fourth columns classify the corresponding symmetry types of the superconducting gap along with their parity properties. In the fifth column, we list the gap structures that are consistent with each symmetry class.}
\label{tab:symmetries}
\centering
\setlength{\tabcolsep}{10pt}
\begin{tabular}{ccccc} 
 \hline\hline
 $M$ & $p$ & Gap symmetry & Parity & Spin state \\ 
 \hline
 0 & 0 & extended $s$ & even & singlet \\ 
 1 & 1 & $p_x + i p_y$ & odd & triplet \\ 
 2 & 0 & $d_{x^2 - y^2} + i d_{xy}$ & even & singlet \\ 
 3 & 1 & $f$ & odd & triplet \\ 
 4 & 0 & $d_{x^2 - y^2} - i d_{xy}$ & even & singlet \\ 
 5 & 1 & $p_x - i p_y$ & odd & triplet \\ 
 \hline\hline
\end{tabular}
\end{table}

As one can see, the model has a similar structure as the Kane-Mele Hamiltonian leading to the opposite Chern number corresponding to the two spin-valleys locked subbands and a conserved time reversal symmetry (TRS) [cf. Fig.\ref{FS1}]. In our parameterization, the phase factors associated with the NNN hopping amplitudes are set to $\phi=2\pi/3$ \cite{wan_1}. Importantly, the orbitals are strongly layer polarized such that the MX (XM) orbital resides predominantly in the top (bottom) layer. This polarization further implies that an applied displacement field $D$ effectively acts as a staggered on-site potential on the honeycomb lattice. As $D$ increases, the relative energy of the two orbitals is shifted which changes the position of the Van Hove singularity as well as the spin-splitting between the spin-up and spin-down Fermi surfaces [cf. Fig. \ref{dos}(b) and Fig. \ref{FS1}]. Also, at at very high vales of the displacement field when the MX and XM orbitals are not mixing anymore, the system undergoes a topological phase transition and becomes topologically trivial.

In our study, we supplement the single particle part of the model described by $\hat{\mathcal{H}}_t$ with the interaction term of the form
\begin{equation}
\hat{\mathcal{H}}_{I} =
U \sum_{i l} \hat{n}_{i l \uparrow}\hat{n}_{i l \downarrow}
+ J_1 \sideset{}{^{\prime}}\sum_{\langle i j\, l  \rangle}
\hat{\mathbf{S}}_{i l} \cdot \hat{\mathbf{S}}_{j \bar{l}} +
J_2 \sideset{}{^{\prime}}\sum_{\langle\langle i j\, l  \rangle\rangle}
\hat{\mathbf{S}}_{i l} \cdot \hat{\mathbf{S}}_{j l}
\label{ham_2}
\end{equation}
where $U$ and $J_{}$ correspond to onsite Coulomb and intersite exchange integrals and the primed summation means that each bond is taken in account only once. The NN exchange coefficient is denoted by $J_1$, whereas the NNN exchange coefficient is denoted by $J_2$. The $J_1$- and $J_2$- terms can lead in a straightforward manner to real space pairing of intersite character within the applied here mean-field treatment as shown below. The superexchange term has been recently studied by us as the origin of the paired state within a single-band approach\cite{zegrodnik_2025,ws_2}.


After applying the Hartree-Fock (HF) approximation to the interaction part, we obtain, 
\begin{equation}
\begin{split}
\hat{\mathcal{H}}_{U} &\approx U \sum_{il\sigma} \Big( \langle{\hat{n}}_{il \sigma}\rangle \hat{n}_{il \bar{\sigma}} - \frac{1}{2}\langle \hat{n}_{il \sigma}\rangle \langle\hat{n}_{il \bar{\sigma}}\rangle\Big) \\
\hat{\mathcal{H}}_{J_1} &\approx
 \frac{-J_1}{2} \sum_{\substack{\langle ijl\rangle \sigma  } } 
 \Big( {{{\Delta}}^{l\bar{l}}_{i j\sigma}} \hat{c}_{i l \sigma}^{\dag} \hat{c}^\dag_{j \bar{l} \bar{\sigma}} + \mathrm{H.c.} + {{{\Delta}}^{l\bar{l}}_{i j{\sigma}}}( {{{\Delta}}^{\bar{l}l}_{ ji{\sigma}}})^{*}\Big) \\
 \hat{\mathcal{H}}_{J_2} &\approx
 \frac{-J_2}{2} \sum_{\substack{\langle\langle ijl\rangle\rangle \sigma  } } 
 \Big( {{{\Delta}}^{l{l}}_{i j{\sigma}}} \hat{c}_{i l \sigma}^{\dag} \hat{c}^{\dag}_{j l \bar{\sigma}} + \mathrm{H.c.} + {{{\Delta}}^{l{l}}_{i j{\sigma}}}( {{{\Delta}}^{{l}l}_{ ji{\sigma}}})^{*}\Big),
\end{split}
\label{ham3}
\end{equation}
 where $\bar{l}=1$ $(2)$ for $l=2$ $(1)$ and $\bar{\sigma}=\uparrow$ ($\downarrow$) for $\sigma=\downarrow$ ($\uparrow$).
\begin{figure*}[t!]
\centering
\includegraphics[width=1\linewidth, height=6.4cm]{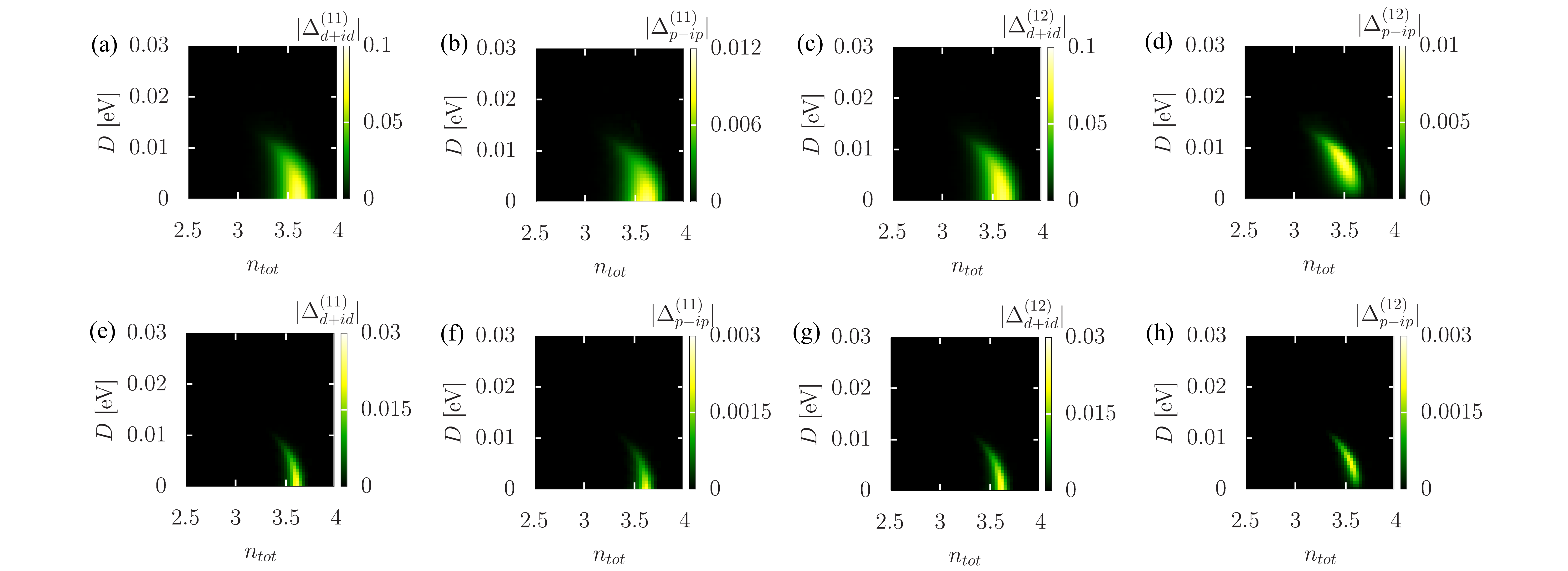}
\caption{ Symmetry resolved superconducting gap amplitudes corresponding to mixed spin-singlet $d+id$ and spin-triplet $p-ip$ paired state as a function of band filling $n$ and displacement field $D$. In (a-d) the exchange interaction parameters are $J_1$ = 5.6 meV and $J_2$ = 2.4 meV, while for the bottom (e-h)  $J_1$ = 2.8 meV and $J_2$ = 1.2 meV. The ratio of the exchange parameters ( $J_1/J_2$ = 7/3) is selected from $4t^2$/$U$ where the on-site Coulomb repulsion $U$ = 95 meV,$|t_{1}|$ = 5.689 meV and $|t_{2}|$ = 3.715 meV. We provide only the $l$ = $1$ component of the gap since the $l$ = $2$ shows the same behavior with approximately same absolute value.}
\label{6_sc}
\end{figure*}

 Exchange-mediated intersite pairing amplitudes of inter- and intra- orbital character are defined as 
\begin{equation}
\begin{split}
  \Delta_{ij\sigma}^{l{l}} &= \langle c_{il \bar{\sigma}} c_{j{l}{\sigma}}\rangle \\
  \Delta_{ij\sigma}^{l\bar{l}} &= \langle c_{il \bar{\sigma}} c_{j\bar{l}{\sigma}}\rangle,
\end{split}
\label{h}
\end{equation}
between site \(i\) of orbital \(l\) and site \(j\) of orbital \(l\) or \(\bar{l}\).
To obtain the momentum-space structure of the gap parameters identified in Eq.~(\ref{h}), the corresponding Fourier transformed corespondents are calculated
\begin{equation}
\begin{aligned}
\Delta_{\mathbf{k}\sigma}^{ll}
&=\sum_{\langle\langle i(j)\rangle\rangle}\Delta_{ij\sigma}^{ll}
e^{i\mathbf{k}\cdot\mathbf{R}_{ij}^{\,ll}},\\
\Delta_{\mathbf{k}\sigma}^{l\bar l}
&=\sum_{\langle i(j)\rangle}\Delta_{ij\sigma}^{l\bar l}
e^{i\mathbf{k}\cdot\mathbf{R}_{ij}^{\,l\bar l}}.
\end{aligned}
\label{kspace}
\end{equation}
Here, $\mathbf{R}_{ij}^{\,ll}$ denotes the six intra-orbital NNN bond vectors, whereas $\mathbf{R}_{ij}^{\,l\bar l}$ denotes the inter-orbital NN bond vectors in which the summation extends over neighbors \( j \) surrounding a central site \( i \).
Both of the NN pairings and NNN pairings with three unique bonds lie entirely within the (\(A_1\oplus E_2\)) subspace of the \(D_{6h}\) point group. But if chirality is present, the pairings encompassing the mixture of $d$- and $p$-wave channels sector will transform completely as the \(E_{2}\) irreducible representation which can be seen easily by applying the \(A_1\) projection operator on the pairing solutions. 

Owing to the sixfold symmetry of the neighboring orbitals, we analyze the symmetry of the superconducting gapped phase by transforming the six independent topological amplitudes, each associated with a specific bond. This allows us to decompose the pairing structure into irreducible components of the underlying lattice symmetry group and thereby identify the corresponding superconducting gap channels, from the following intra- and inter-orbital gap channels,
\begin{equation}
\begin{split}
    \Delta^{ll}_{M,p } &= \frac{i^p}{6}\sum_{i(j)}e^{-iM\varphi_{ji}}\Delta_{ij\sigma}^{l{l}}, \\
    \Delta^{12}_{M,p} &= \frac{i^p}{6}\sum_{i(j)l}e^{-iM\theta_{ji}}\Delta_{ij\sigma}^{l\bar{l}},
\end{split}
\label{realspace}
\end{equation}
 where the angle \( \varphi_{ji} \) and \( \theta_{ji} \) denotes the orientation between the vector \( \mathbf{R}_{ji} = \mathbf{R}_j - \mathbf{R}_i \) and the positive \( x \)-axis. 
For the intra-orbital channel we have six NNN $j$ sites for each of the two $l$ orbitals with angles taking the discrete values  \( \left\{ 0, \pi/3, \pi, 2\pi/3, 4\pi/3, 5\pi/3 \right\} \) which capture the angular positions of neighbors in a hexagonal arrangement. For the inter-orbital channel there are three NN $j$ sites for $l=1$ and another three for $l=2$ which together form a set of six different angles  \( \left\{ \pi/6, \pi/2, 5\pi/6, 7\pi/6, 3\pi/2, 11\pi/6 \right\} \). The parameter \( M \) labels the symmetry channel and is restricted to integers, each specifying a distinct pairing configuration among the six allowed symmetries (see Table 1). Meanwhile, the parity index \( p \) differentiates between even and odd symmetries, taking the value \( p = 0 \) for even and \( p = 1 \) for odd parity cases.

\section{\label{sec:level3}Results}
We now characterize the unconventional superconducting state within the HF mean-field description of an effective two band model of TMD homobilayer. We focus on analyzing the stability of superconducting phase as a function of band filling and displacement field together with studying the gap symmetry as well as its topological features.

\subsection{Paired state features}\textbf{}
First, we determine the stability range and symmetry of the obtained superconducting amplitudes across the ($n$,$D$) diagram.
In Fig.\ref{6_sc}, we provide the results calculated for two selected values of the exchange interaction $J_1$ and $J_2$. Our approach leads to stability of the mixed $d+id$ (singlet) and $p-ip$ (triplet) paired state. Note that we distinguish between intra- and inter-orbital SC gap amplitudes, which originate from the $J_1$- and $J_2$-terms of our Hamiltonian [cf. Eq. (\ref{ham_2})], respectively. As one can see all types of gap parameters (singlet/triplet, intra-/inter- orbital) become non-zero in the same ranges of parameters and for relatively significant values of the pairing strengths ($J_1$ and $J_2$) the SC stability spreads across a substantial part of the phase diagram [cf. Fig.\ref{6_sc} (a-d)]. However as expected, by decreasing the values of $J_1$ and $J_2$ one reduces the paired state stability region as shown in Fig.\ref{6_sc} (e-h).

To estimate a realistic balance between the intra- and inter-orbital pairings, the value $J_1/J_2$, is determined by the use the relations: $J_1 = 4|t_1|^2/U$ and $J_2 = 4|t_2|^2/U$, where $U$ represents the Coulomb repulsion term.
In practice, modifying the twist angle effectively changes $U$ which allows to tune the strength of correlations.
Moreover, the dielectric constant can be altered by the three-dimensional dielectric environment which significantly affects $U$. For the TMD homobilayers with few angle twist it is estimated that $U$ is roughly comparable to $W$.

In our study we estimate the onsite Coulomb repulsion from the standard Gaussian Wannier approximation as $U \approx e^2\sqrt{\pi}/(4\pi\epsilon_0\epsilon_{\mathrm{eff}}l)$. By taking $l \approx 3$  nm as the extent of the maximally localized Wannier orbital \cite{wan_2} for the  moir\'{e} lattice constant $a_M \approx 5.2$ nm and $\epsilon_{\mathrm{eff}} \approx 8\text{--}10$ (for an hBN-encapsulated device with metallic gates), we obtain $U \approx 85\text{--}105$ meV. Throughout this work, we take $U = 95$ meV unless specified otherwise. Such value is also within the estimated range resulting from the projection procedure after Wannerization for the twisted TMD bilayers \cite{wan_1}.

\begin{figure}[b]
\centering
\includegraphics[width=0.975\linewidth, height=6.9cm]{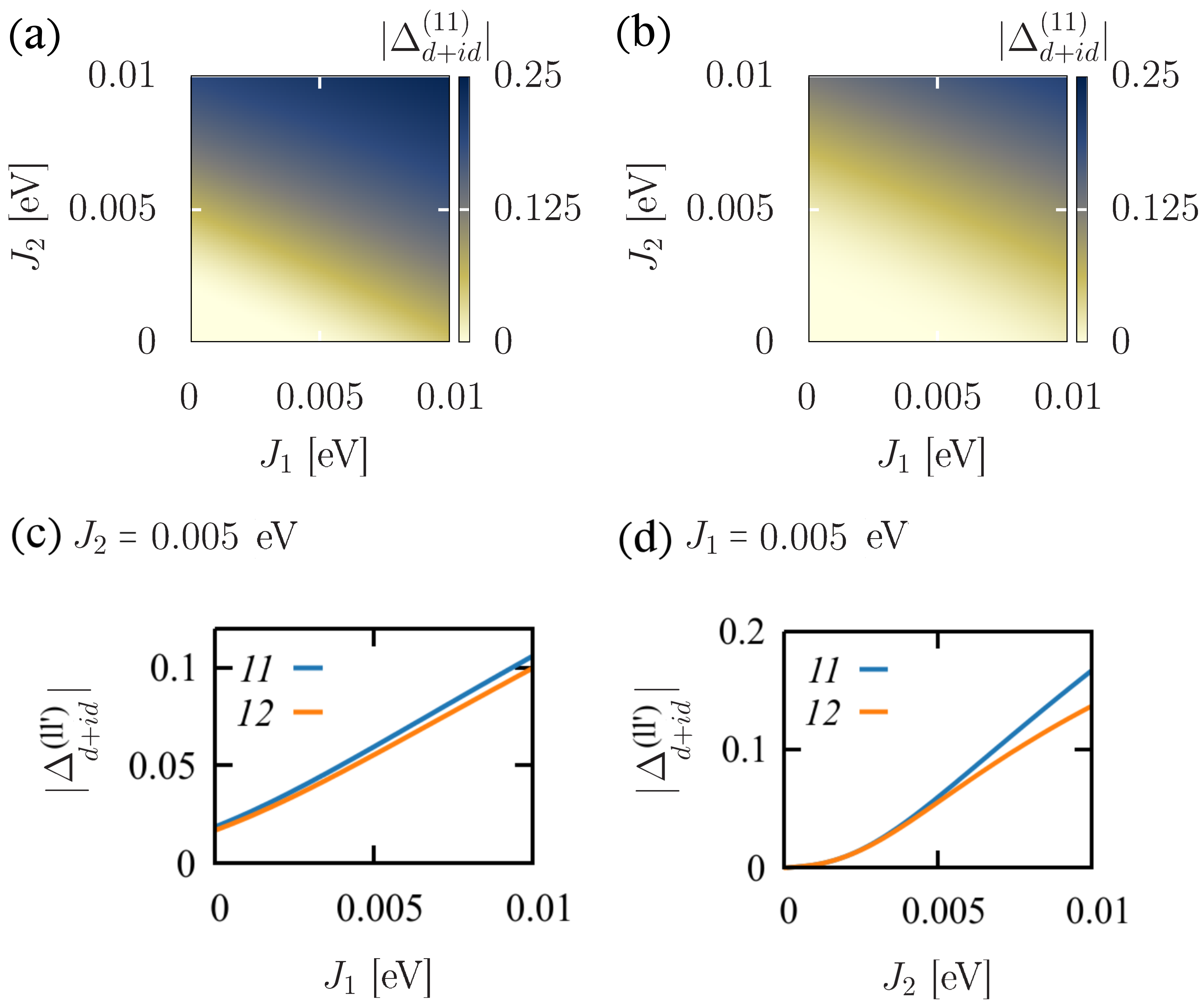}
\caption{Superconducting singlet gap at band filling $n_{\mathrm{tot}}=3.5$ as a function of $J_1$ and $J_2$ for (a) $D=0$ and (b) $D=10$ meV. For $D=10$ meV, panels (c) and (d) show the induced inter-orbital and intra-orbital gap components upon varying $J_1$ at fixed $J_2=5$ meV and varying $J_2$ at fixed $J_1=5$ meV, respectively. The triplet component exhibits the same qualitative evolution and is not shown for clarity.}
\label{J1J2}
\end{figure}

\begin{figure*}[t!]
\centering
\includegraphics[width=0.825\linewidth, height=8.75cm]{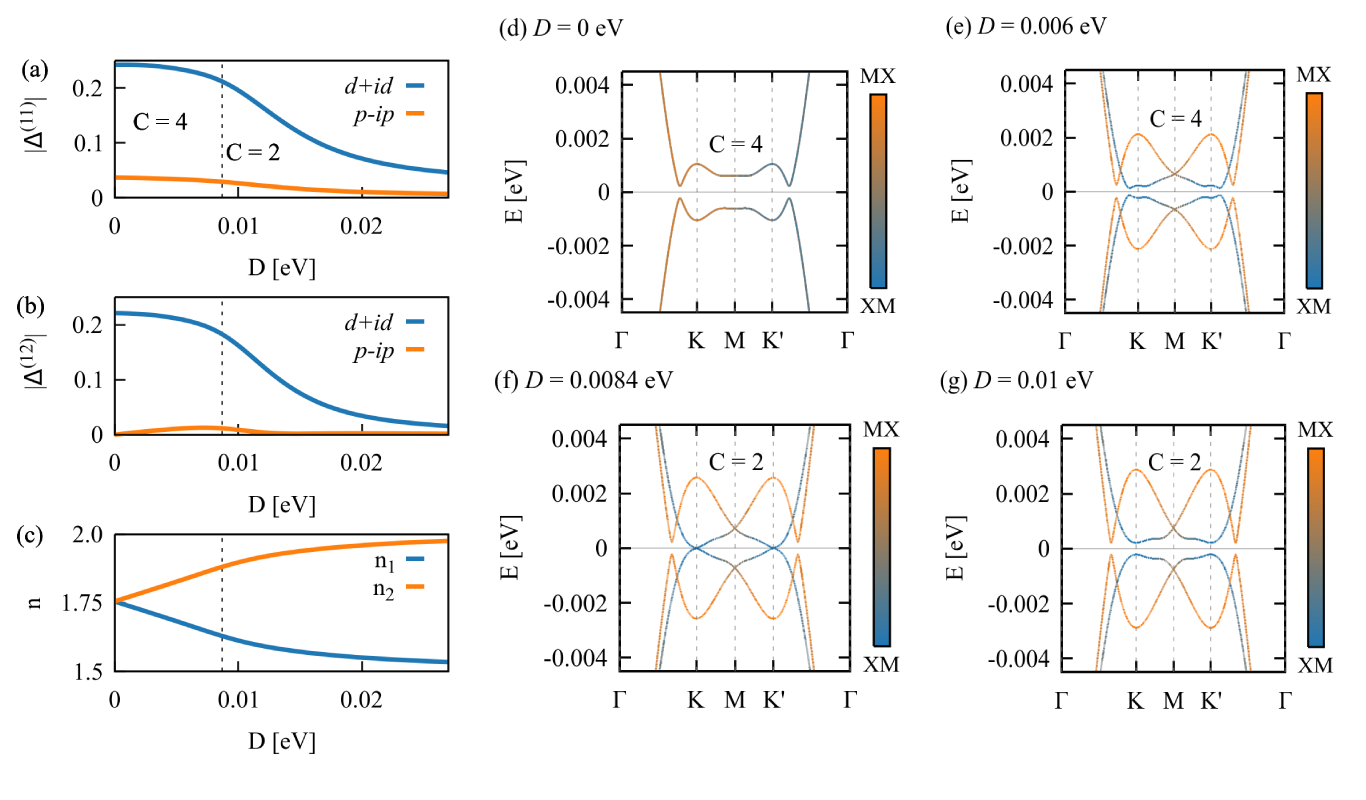}
\caption{Panels (a) and (b) show the magnitude of the superconducting gap for the mixed $d+id$ and $p-ip$ pairing states, while panel (c) presents the orbital-resolved carrier concentration as a function of $D$ for $n_{\mathrm{tot}}=3.5$, $J_1=J_2=10\,\mathrm{meV}$, and $U=95\,\mathrm{meV}$. The dashed line marks the topological transition in the superconducting bands. Panels (d)--(g) show the corresponding superconducting band structures for increasing displacement field. At $D=0$ in (d), the upper superconducting bands are degenerate and carry a total Chern number $C=4$; at $D=6\,\mathrm{meV}$ in (e), the degeneracy is lifted while the system remains in the same $C=4$ phase. Near $D=8.4\,\mathrm{meV}$ in (f), the relevant bands touch at the topological transition, whereas at $D=10\,\mathrm{meV}$ in (g) the gap reopens and the total Chern number changes to $C=2$. }

\label{bands2}
\end{figure*}

As one can see from Fig. \ref{6_sc}, our calculations show that both singlet- and triplet-gap amplitudes behave in a similar manner, although the triplet component is roughly an order of magnitude smaller than the singlet corespondent.
In the considered situation, intrinsic SO coupling gives rise to singlet-triplet mixing. In the absence of the SO coupling, the pairing is only of spin-singlet character. Also, the stability regime of the paired state is concentrated around the area which represents increased density of state due to the van Hove singularity evolution on the phase diagram [cf. Fig. \ref{dos}(b)]. This is due to the fact that increased density of states at the Fermi level stabilizes the superconducting state. Nevertheless, with increasing displacement field the pairing is relatively quickly suppressed. 

It should be noted that for the $D=0$ case the intra-orbital triplet component is already finite, while some non-zero displacement field is necessary to induce triplet pairing within the inter-orbital channel. From the technical point of view, this is due to the fact that the intra-orbital hoppings posses a direction and spin dependent complex phase which already reduce the $C_6$ symmetry of the lattice to $C_3$ symmetry within the two orbital sectors separately. As previously shown for the case of a single band description of tWSe$_2$ such symmetry reduction leads to a singlet-triplet mixing \cite{ws_1,ws_2}. Nevertheless, the inter-orbital pairing is purely real in the considered approach, therefore, the actual appearance of the triplet inter-orbital pairing requires an additional symmetry breaking or band structure effect supplied by some finite value of $D$.


Next, we analyze how the balance between the two types of exchange interaction strengths influences the paired state. In Figs. \ref{J1J2} (a) and (b) we show that, by increasing both $J_1$ and $J_2$, the pairing amplitudes become stronger for two selected values of the displacement field: $D$ = $0$ and $D$ = 10 meV. 
For any point at the ($J_1$-$J_2$) plane, the NNN pairing channel always dominates over the NN pairing channel.  
As one can see from Fig.\ref{6_sc} the region of inter-orbital singlet gap amplitude completely overlaps with the intra-orbital singlet gap amplitude. It is also the case here, and thus we only show the evolution of the intra-orbital pairing channel for clarity. It should be noted that even for the case of purely intra-orbital exchange scenario when $J_{1}=0$, the inter-orbital gap is still non-zero, while it is not the case in the reversed situation as shown explicitly in Fig. \ref{J1J2}(c,d). This is due to the fact that the intra-orbital pairing together with the single-particle hybridization creates an effective pairing between the orbitals.


\subsection{Topological properties of the paired state}

Here we analyze the non-trivial topology of the obtained paired state and possible topological phase transitions induced by the displacement field. In this part of our study we set significant exchange interactions strength $J_1$ = $J_2$ = 10 meV so that the superconducting gap is clearly visible in the band structure for the considered range of $D$ values. In Fig. \label{bands2} (a-b) we show how the SC gap amplitudes are being suppressed with increasing displacement field for a selected value of band filling $n_{tot}=3.5$. This effect is again due to the Van Hove singularity evolution. Namely, for this specific value of band filling, the density of states at the Fermi level is decreasing with increasing displacement field, which gradually weakens the pairing [cf. Fig. \ref{dos}(b)]. Additionally, as shown in \label{bands2}(c), the displacement field also tunes the relative balance between the occupation of the two orbitals of the model. In Fig. \ref{bands2}(d-g) we show the quasiparticle dispersion relations together with the contribution resulting from the MX ($l=1$) and XM ($l=2$) orbitals marked by the colored scale. As can be seen, both MX and XM states contribute to the SC gap opening for the selected $D$ values. It should be noted that for relatively high values of the displacement field ($D\approx0.02$\;eV) the XM ($l=2$) states become almost completely filled [cf. Fig. \ref{bands2}(c)] and therefore the gap opens in the band dominated by the MX ($l=1$) states. For completeness we show this effect in Appendix B.

Interestingly, according to our analysis the obtained paired state is topologically non-trivial since the calculated Chern number takes a non-zero value. Moreover, a topological phase transition appears at some critical value of $D\approx 0.0084\;$eV where the Chern number changes from $C=4$ to $C=2$. In Fig. \ref{bands2}(d-g), we show the calculated band structure for four selected values of $D$. As $D$ becomes non-zero, first, the degeneracy is lifted, and two spin-split underlying Fermi surfaces emerge, both of which are fully gapped. However, at the critical value of $D$ the gap closes in the proximity of the $K$ and $K'$ points of the Brillouin zone, which marks the topological phase transition. Since the Chern number is a topological invariant it can change its value only via such gap closing and reopening process.

In interpreting Figs. \label{bands2} (d-g), it is important to note that the Chern number is the sum of all the topological charges contained inside the Fermi surface. For the case of paired state, each topological charge corresponds to a nodal point in $\mathbf{k}$-space at which the SC gap closes \cite{ws_2}. Since the displacement field affects the shape of the Fermi surfaces [as seen in Fig. \ref{FS1}(c,d)], at a critical value of $D$, some of the nodal points may be moved outside the Fermi surface, leading to a topological transition. Exactly at the critical value of $D$ the Fermi surface crosses the nodal points resulting in a gap closing as shown in Fig. \ref{bands2}(f). For the sake of completeness in Appendix A we show the intra- and inter- orbital $\mathbf{k}$-dependent gap amplitudes corresponding to the obtained mixed $d+id$/$p-ip$ state.

\section{Conclusions}
In summary we carry out a mean-field investigation of the paired state in an effective model of twisted moir\'e TMD bilayer. 
We focus on the two top moir\'{e} valence bands described by a two orbital Kane-Mele model supplemented with an onsite Coulomb repulsion and an intersite exchange terms responsible for the pairing mechanism. Since the model resembles the main features of the Kane-Mele Hamiltonian, such approach allows us to study the principal features of the real-space paired state formed in an environment created by topologically non-trivial bands. 

In the proximity of band filling of $n_{tot}=3.5$, we find the emergence of a singlet-triplet mixed SC state mostly dominated by the singlet pairing for both intra-
and inter-orbital pairing channels. Within our mean-field approach the stability range of the paired state is determined by the evolution of the van Hove singularity across the $(n_{tot},D)$-phase diagram. The singlet-triplet mixing obtained here is a consequence of SO coupling. Although the complex nature of the pairing already reduces the gap symmetry from $C_6$ to $C_3$, it does not by itself guarantees a finite inter-orbital triplet contribution. Finite displacement field lifts the degeneracies thereby allowing the symmetry permitted ($p-ip$) interorbital component to become non-zero.

We also calculate the Chern number using the BZ triangulation method in order to extract the general topological features of the paired state. For the selected band filling, the obtained total Chern number of the paired state is $C$ =$\pm $4 within the regime of low displacement fields. However, with increasing $D$ a topological phase transition occurs which is marked by a SC gap closing and reopening at which the Chern number changes its value from $C$ =$\pm $4 to $C$ =$\pm$ 2. This effect can be explained based on the changes of the relative position of the underlying Fermi surfaces and topological charges induced by the displacement field.

It should be noted that in the twisted TMD bilayers apart from superconductivity also other symmetry broken and topologically non-trivial states are observed in the phase diagram. Namely, for twisted WSe$_2$ antiferromagnetism, Mott insulating behavior, and a series of QAH states have been discussed \cite{Guo2026,Feldman2024,Xia2026}. For the case of twisted MoTe$_2$ bilayer magnetic and charge ordering as well as various topologically non-trivial states like QAH and FQAH have been identified experimentally \cite{exp_sc,sun2026twistangleevolutionvalleypolarizedfractional}. To better determine the stability range of the superconducting phase analyzed theoretically here one should in principal take into account also the possibility of those other states since the interplay and/or competition between different tendencies can alter the area of the phase diagram corresponding to the paired state. Also, in order to reconstruct the typical strong correlation induced states such as Mott insulators as well as the influence of band renormalization on the pairing one would have to go above the mean-field treatment applied here. Incorporating all the mentioned effects should be considered as a highly complex task even within the effective two-band approach. Such analysis is beyond the scope of this work.

Here, a final remark is in order. Namely, the paired state resulting from our approach develops from valley-degenerate Fermi surfaces, which is consistent with the recent experimental findings related with the observation of  superconductivity in tWSe$_2$ for twist angles between $3.6^{\circ}$ to $5^{\circ}$ as well as in tMoTe$_2$ for the twist angle of $5.78^{\circ}$ \cite{sun2026twistangleevolutionvalleypolarizedfractional}. In such a case the pairing most likely takes place between two different spins (valleys), which correspond to singlet and/or triplet (with $S_z=0$) pairing. Such a situation should be distinguished from the triplet $S_z=\pm 1$ case which is consistent with the paired state developed from a valley-polarized Fermi surface exhibiting an anomalous Hall response as observed for the case of tMoTe$_2$ for $\theta=3.83^{\circ}$ \cite{exp_sc}. 

The data evaluated from the numerical calculations are available in the open repository \cite{zegrodnik_michal_2026_7}.

\begin{acknowledgments}
 This  research was partly supported by National Science Centre, Poland (NCN) according to Decision No. 2021/42/E/ST3/00128 and partly by program “Excellence initiative–research university” for the AGH University of Krakow. 

\appendix
\section{Momentum dependence of the superconducting gaps}
For the sake of completeness, we also show explicitly the \textbf{k}-dependence of the superconducting gap amplitudes in Fig. \ref{nodal} for selected values of the model parameters.
As one can see in (a,b) for the intra-orbital gap, apart from the nodal points with topological charge being $c=-1$ at the high symmetry points, there are three additional ones, each corresponding to the topological charge equal $c=1$ in the close proximity of the \(\Gamma\) point. 
\begin{figure}[h!]
\centering
\includegraphics[width=0.9\linewidth, height=8.25cm]{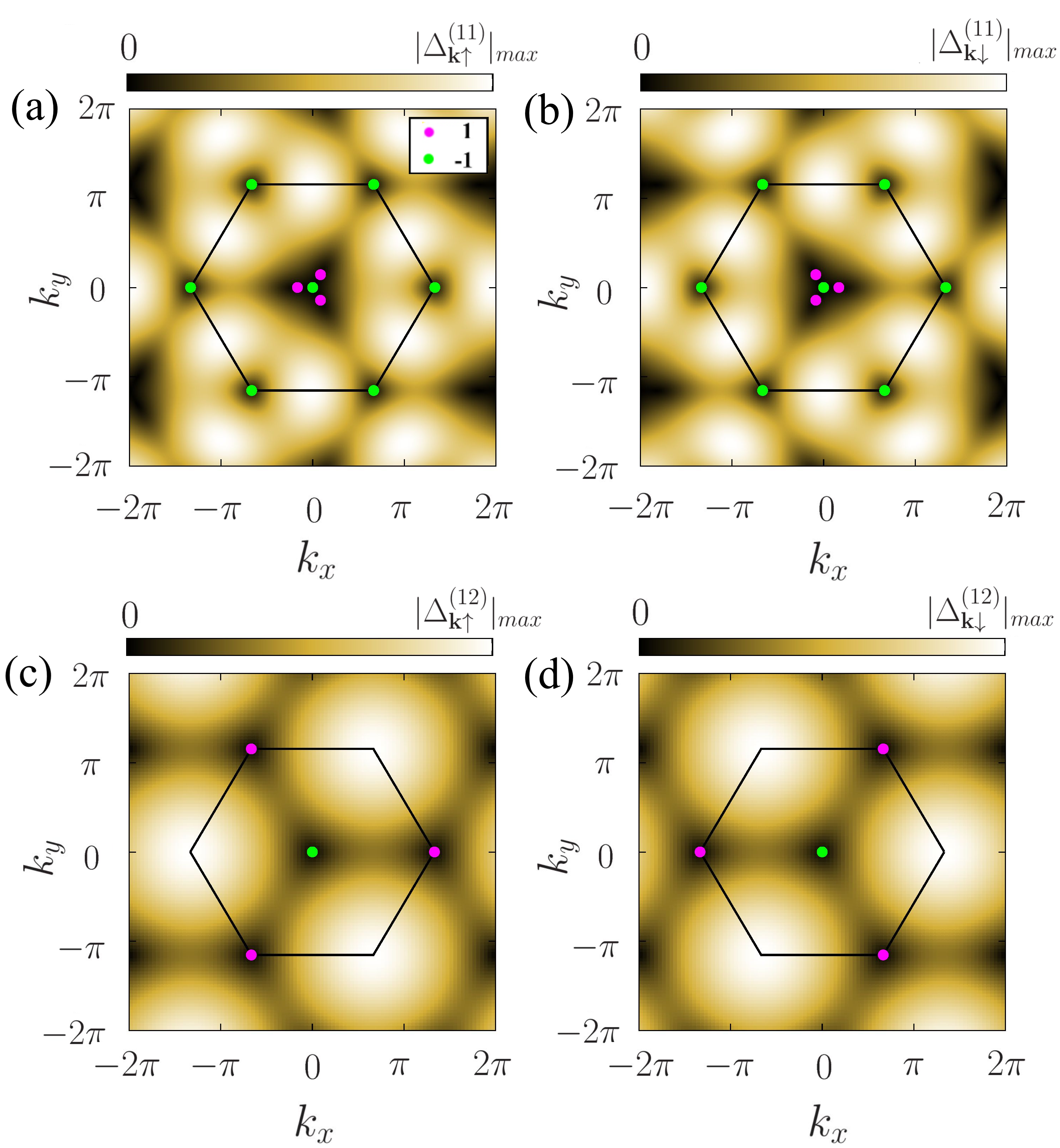}
\caption{Absolute value of the superconducting gap 
as a function of momentum (\textbf{k}) for the mixed singlet-triplet channel when $J_1=J_2=10$\;meV, $D=10$\;meV and $U=95$\;meV. The magenta and green dots mark the positions of the Chern charges of the value $1$ and $-1$, respectively. (a,b) Intra-orbital pairing gap amplitudes, whose nodal points form a \(C_6\)-symmetric pattern governed by the neighboring hoppings, while (c,d) Inter-orbital pairing gap amplitudes, exhibits a \(C_3\)-symmetric \((2\pi/3\)-rotated) pattern associated with the inter-orbital site configuration.}
\label{nodal}
\end{figure}
All the values of the topological charges provided here were determined by calculating the winding of the $\mathbf{k}$-dependent superconducting gap complex phase around them. Note that such structure of the nodal points resemble the $\mathbf{k}$ dependent gap in the single band description of the tWSe$_2$ studied earlier \cite{ws_2}. In both situations the gap structure corresponds to a mixture of $d+id$ and $p-ip$ symmetry. That is because the intra-orbital hoppings and exchange interactions for $l=1$ and $l=2$ separately resemble the single band model scenario for the case of nonzero displacement field. However, here we also have the inter-orbital pairings which do not have any single band corespondents. As we show for the inter-orbital case the pairing follows a $2\pi/3$ rotated patterns corresponding to the interorbital NN sites. It should be noted that due to the single particle hybridization terms, which mix the states originating from the two orbitals, the $\mathbf{k}$-dependence of the intra- and inter- orbital gaps do not completely determine the value of the SC gap which is opening around the Fermi level. Instead, the resulting SC gap which opens around the Fermi level is a combination of all the $\Delta^{ll'}_{\mathbf{k}\sigma}$'s as well as the single particle hybridization term.

\section{Superconducting gap opening in the high-$D$ regime}
For the sake of completeness, here we show the superconducting band structure of the considered system in the regime of relatively high values of the displacement field. As seen from Eq.~(\ref{ham1}) the displacement field generates a staggered potential which results in charge transfer from the MX ($l=1$) to the XM ($l=2$) lattice sites [cf. Fig. (c)]. As a consequence for high enough $D$ the XM states become nearly doubly occupied and the band, which is crossing the Fermi energy is dominated by the MX states. The system becomes more and more similar to a triangular lattice with the MX orbitals while the second triangular lattice with the XM orbitals is fully filled and therefore does not contribute to the physical features of the system. In Fig. \ref{band_high_D} we show the superconducting band structure for relatively high value of the displacement field ($D=0.02$\;eV) where it is seen that mostly the MX states contribute to the superconducting gap opening.

\begin{figure}[h!]
\centering
\includegraphics[width=0.9\linewidth]{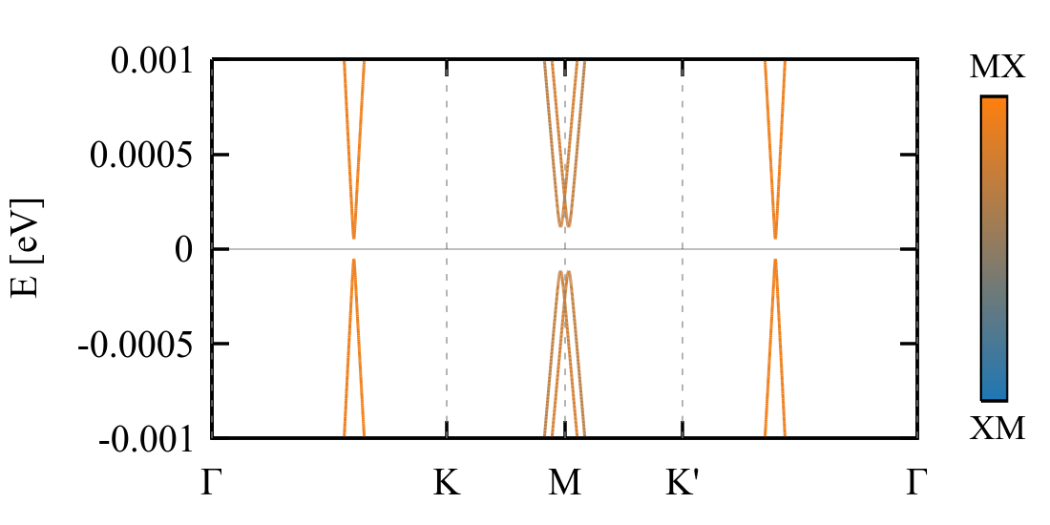}
\caption{The superconducting band structure for the case of relatively high value of displacement field, $D=0.02$\;eV. All the other model parameters are the same as in Fig. \ref{bands2}(d-g). Note that in the high displacement field regime the XM ($l=2$) states are nearly doubly occupied ($n_2\approx2$), therefore the SC gap opening is mostly due to the pairing between the MX states.}
\label{band_high_D}
\end{figure}

\end{acknowledgments}

\bibliography{refs}

\end{document}